\documentclass[prl,twocolumn,reprint,amsmath,amssymb,superscriptaddress,floatfix,longbibliography]{revtex4-1}
\usepackage[breaklinks=true,colorlinks,citecolor=blue,linkcolor=blue,urlcolor=blue]{hyperref}
\usepackage{float}
\usepackage{epsfig,mathrsfs,color,latexsym,subfigure,marginnote}

\def\be{\begin{equation}}
\def\ee{\end{equation}}

\def\bea{\begin{eqnarray}}
\def\eea{\end{eqnarray}}

\begin{document}
\title{Low-energy type-II Dirac fermions and spin-polarized topological surface states in transition-metal dichalcogenide NiTe$_2$ }

\author{Barun Ghosh \footnote{These authors contributed equally to this work.}}
\affiliation{Department of Physics, Indian Institute of Technology - Kanpur, Kanpur 208016, India}

\author{Debashis Mondal}
\affiliation{Istituto Officina dei Materiali (IOM)-CNR, Laboratorio TASC, in Area Science Park, S.S.14, Km 163.5, I-34149 Trieste, Italy.}

\author{Chia-Nung Kuo}
\affiliation{Department of Physics, National Cheng Kung University, 1 Ta-Hsueh Road 70101 Tainan, Taiwan}

\author{Chin Shan Lue}
\affiliation{Department of Physics, National Cheng Kung University, 1 Ta-Hsueh Road 70101 Tainan, Taiwan}

\author{Jayita Nayak}
\affiliation{Department of Physics, Indian Institute of Technology - Kanpur, Kanpur 208016, India}

\author {Jun Fujii}
\affiliation{Istituto Officina dei Materiali (IOM)-CNR, Laboratorio TASC, in Area Science Park, S.S.14, Km 163.5, I-34149 Trieste, Italy.}

\author{Ivana Vobornik}
\email{ivana.vobornik@elettra.trieste.it}
\affiliation{Istituto Officina dei Materiali (IOM)-CNR, Laboratorio TASC, in Area Science Park, S.S.14, Km 163.5, I-34149 Trieste, Italy.} 

\author{Antonio Politano}
\email{antonio.politano@iit.it}
\affiliation{Dipartimento di Scienze Fisiche e Chimiche (DSFC), Universit\`a dell'Aquila, Via Vetoio 10, I-67100 L'Aquila, Italy}

\author{Amit Agarwal} 
\email{amitag@iitk.ac.in}	
\affiliation{Department of Physics, Indian Institute of Technology - Kanpur, Kanpur 208016, India}

\begin{abstract} 
Using spin- and angle- resolved photoemission spectroscopy (spin-ARPES) together with ${\it ab~initio}$
calculations, we demonstrate the existence of a type-II Dirac
semimetal state in NiTe$_2$. We show that, unlike PtTe$_2$,
PtSe$_2$, and PdTe$_2$, the Dirac node in NiTe$_2$ is located in close 
vicinity of the Fermi energy. Additionally, NiTe$_2$ also hosts a
pair of band inversions below the Fermi level along the $\Gamma-A$ high-symmetry direction, with one of them leading to a Dirac cone in the surface
states. The bulk Dirac nodes and the ladder of band inversions in
NiTe$_2$ support unique topological surface states with chiral
spin texture over a wide range of energies. Our work paves the way
for the exploitation of the low-energy type-II Dirac fermions in
NiTe$_2$ in the fields of spintronics, THz plasmonics and
ultrafast optoelectronics.


\end{abstract}

%
%
\maketitle

%
The discovery of topological semimetals has ushered in a new era of exploration of massless relativistic quasi-particles in crystalline solids\cite{Bansil2016,Hasan2010,Ashvin2018, kong11}. These arise as emergent quasi-particles in crystals with linearly dispersing bands in vicinity of a degenerate band crossing point (either accidental or symmetry-enforced) and are protected by crystalline symmetries\cite{bradlyn16}. Double, triple and
quadruple degeneracy of the band crossing leads to topologically
protected Weyl\cite{xu15a,huang15,lv15,xu15b,fang12,weng15a,bahadur12,PhysRevLett.119.026404,Xue1603266},
triple point\cite{lv17,gao18,zhu17,lv17,yu17,fulga17,yang17,wang17} and Dirac fermions\cite{young12,young15,wieder16,liu14a,liu14b,neupane14,borisenko14,wang12,wang13},
respectively. In contrast to their high energy counter-parts, these emergent quasi-particles are not 
protected by Lorentz symmetry, and can also occur in a tilted form, giving rise to type-I and type-II Dirac fermions. 
Specifically,  Na$_3$Bi\cite{wang12,liu14b} and Cd$_3$As$_2$\cite{wang13,liu14a} are type-I Dirac semimetal (DSM), while the transition-metal dichalcogenides (TMDs) PtTe$_2$\cite{PtTe2_natcomm,Bahramy2017,PtTe2PRL}, PtSe$_2$\cite{PtSe2_PRB}, and PdTe$_2$\cite{PdTe2_PRL,Bahramy2017,PhysRevB.96.041201} are type-II DSM. 
\begin{figure}[h!]
\includegraphics[width=0.95\linewidth]{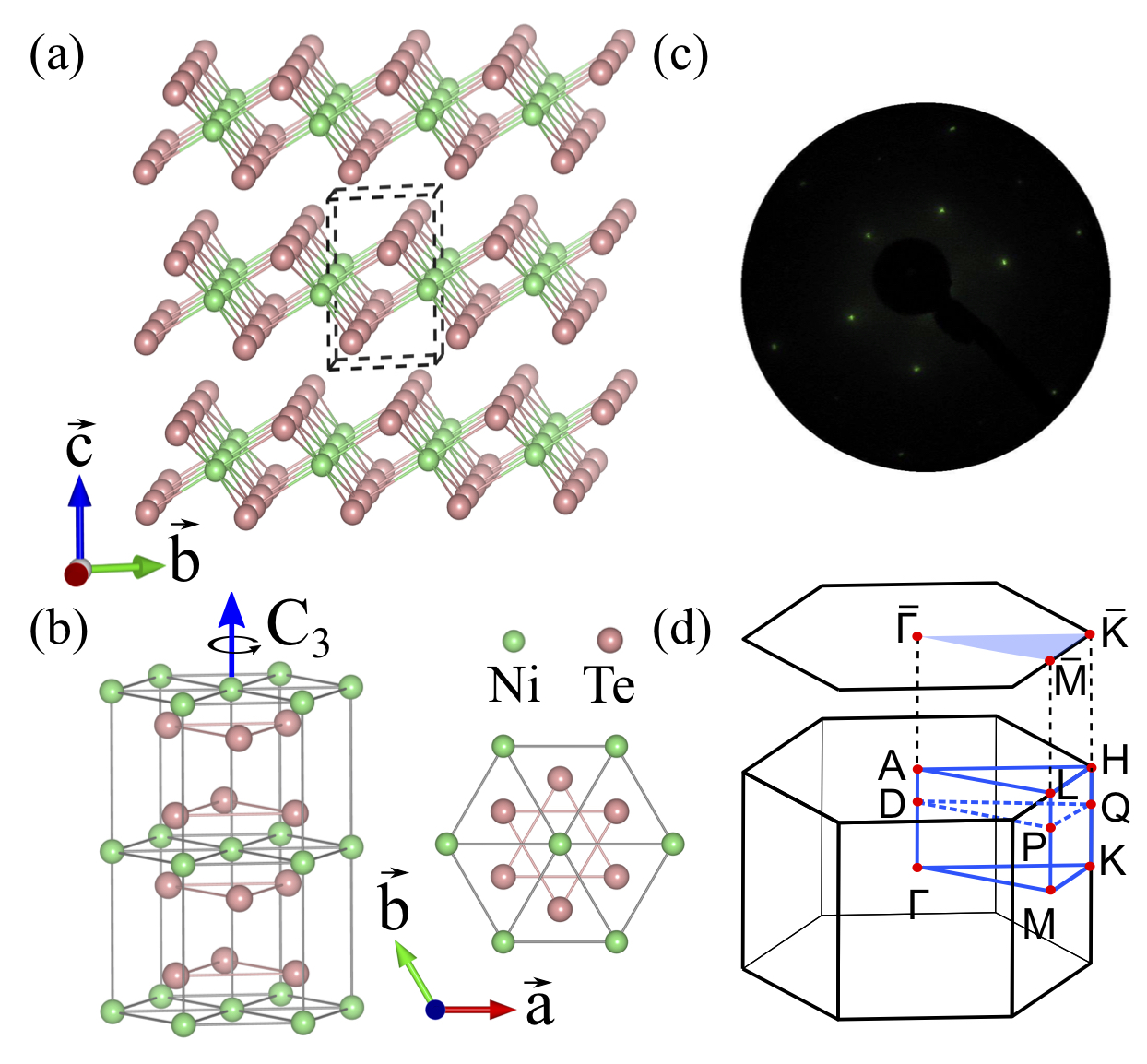}
\caption{(a) The side view and (b) hexagonal crystal structure of NiTe$_2$ with the $C_3$ rotation axis. Layers of Ni are stuffed between two Te layers. 
(c)  The LEED pattern of (0001)-oriented  NiTe$_2$ single crystals, acquired at a primary electron beam energy of 84 eV, clearly indicates its purity and the six-fold symmetry along along the (001) direction. (d) The bulk and the (001) surface Briluoin zone (BZ) of NiTe$_2$. 
\label{fig1}}
\end{figure}
\begin{figure*}
\includegraphics[width=0.9\linewidth]{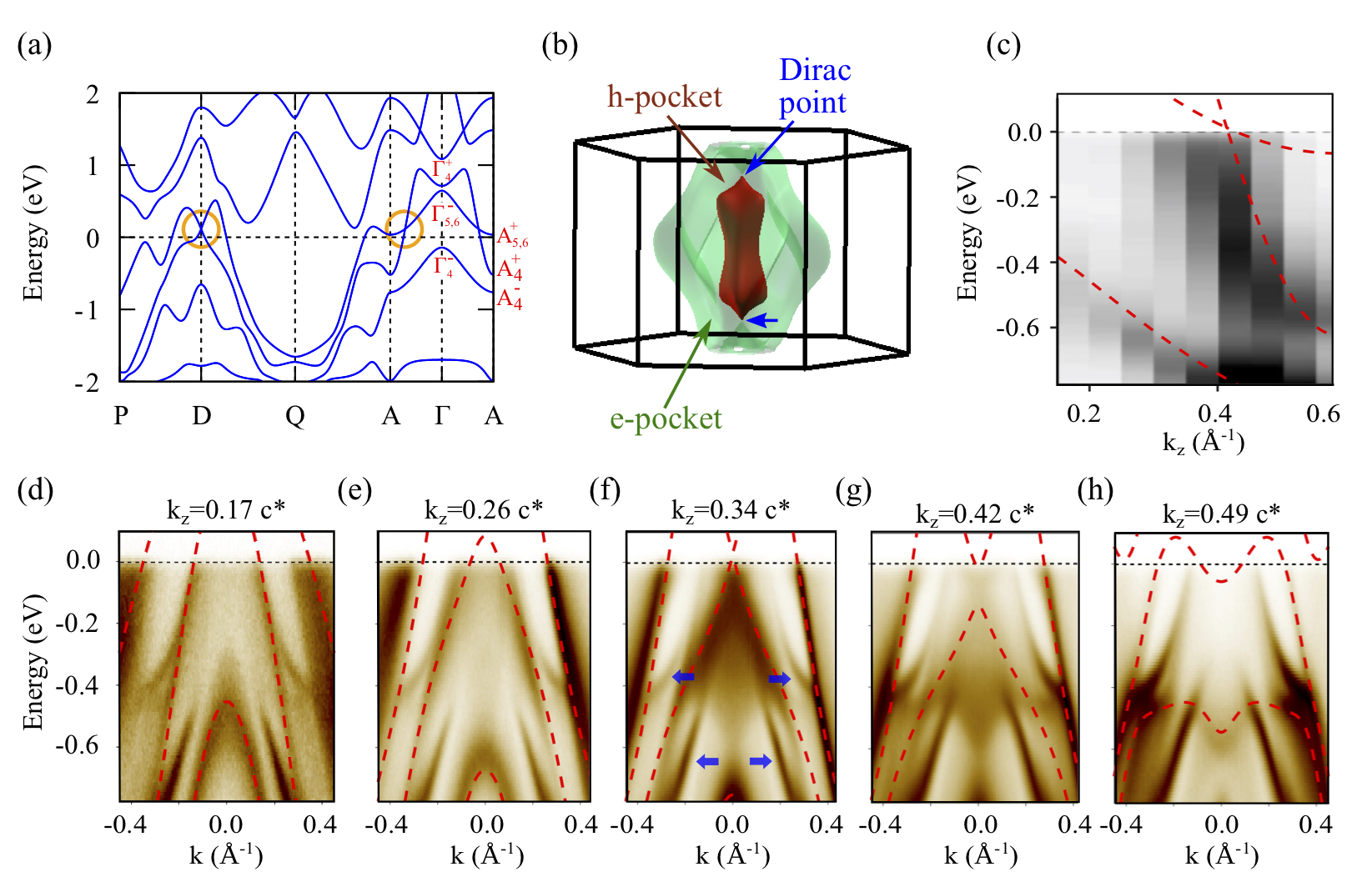}
\caption{(a) Band structure of NiTe$_2$ (including SOC)  clearly showing the tilted type-II Dirac node along the $\Gamma-A$ direction (at D). 
The irreducible representation of the bands  (close to Fermi energy) at the $\Gamma$ and $A$ points are also marked. (b) The Fermi surface of NiTe$_2$ originating from the crossing Dirac bands. 
The type-II Dirac points appear at the touching points of the electron and hole pockets. (c) $k_z$ dispersion along $\Gamma-A$ deduced from the $h\nu$-dependent data measured along ${\bar \Gamma}-{\bar K}$ direction (shown in part in (d-h)); red dashed lines represent the DFT calculations; 
(d)-(h) The measured band dispersion along the ${\bar K}-{\bar \Gamma}-{\bar K}$ direction for different $k_z$ values, with the red dashed lines indicating the bulk DFT band structure. The blue arrows in panel (f) mark the surface states. Data were taken at photon energies of 17, 19, 21, 23 and 25 eV, respectively. Panel (f) corresponding to $k_z=0.34 ~c^*$, is closest to the location of the Dirac point ($k_z = 0.35~c^*$ in our DFT calculations). Note that, for matching with the experimental data, we have shifted the DFT band structure 
downward by 100~meV.}
\label{fig2}
\end{figure*}

In group X Pd- and Pt- based dichalcogenides, the bulk Dirac node lies deep below the Fermi level ($\sim$0.6, $\sim$0.8 and $\sim$1.2 eV in PdTe$_2$, PtTe$_2$, and PtSe$_2$, respectively) \cite{PtTe2_natcomm,Bahramy2017,PtTe2PRL,PtSe2_PRB,PdTe2_PRL,Bahramy2017,PhysRevB.96.041201}, hindering their successful exploitation in technology. 
%
%
%
In contrast, NiTe$_2$ has been predicted to host type-II Dirac fermions in vicinity of the Fermi energy \cite{doi:10.1021/acs.chemmater.8b02132}. The so far performed experimental studies on NiTe$_2$ have primarily focused on its crystal structure, and transport properties while its topological band structure remains unexplored \cite{doi:10.1143/JPSJ.11.21, ETTENBERG1970583,Guo_1986,Orders_1982,MONTEIRO2017129,doi:10.1021/acs.chemmater.8b02132,doi:10.1021/jacs.8b08124,PhysRevB.99.155119}. Motivated by this, we explored the electronic band structure of NiTe$_2$ by means of spin- and angle-resolved photoemission spectroscopy (ARPES) in combination with density functional theory (DFT).


\begin{figure*}
\includegraphics[width=0.9\linewidth]{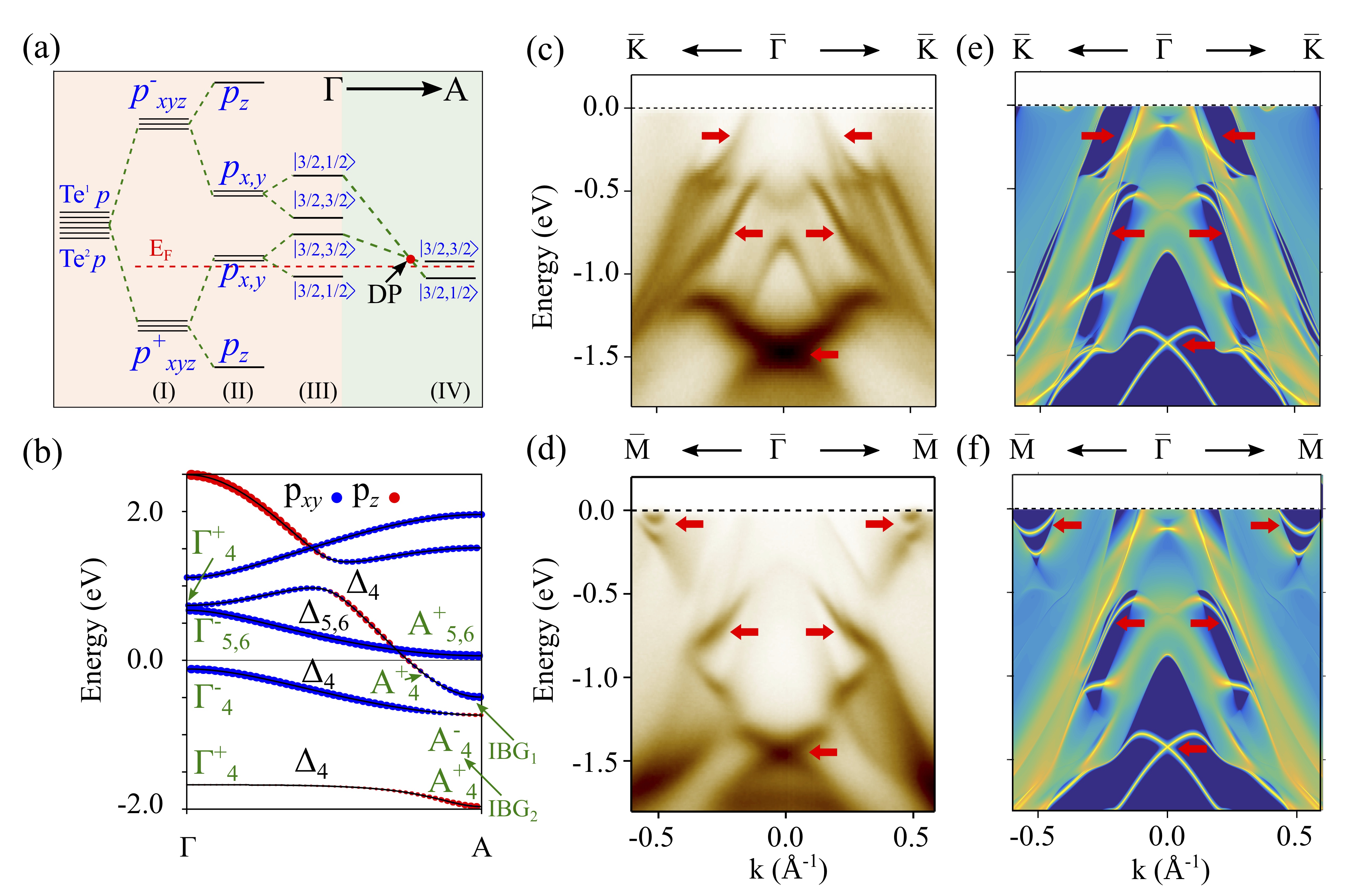}
\caption{(a) The evolution of the Te 5p orbitals in the formation of Dirac-cone states in NiTe$_2$. Step (I) shows the the creation of bonding and anti-bonding orbitals. Step (II) shows the effect of the strong trigonal crystal field which separates the $p_z$ orbitals from the $p_{x,y}$ orbitals. In step (III), we show the splitting of these states into the $|J,|m_J|\rangle$ states in the presence of SOC. In step (IV), we demonstrate the effect of out of plane dispersion and the formation of the Dirac point. (b) The 
orbital-resolved band structure and various band inversions along the $\Gamma A$ high-symmetry direction is shown along with the irreducible representations of the bands. In panels (c) and (d) we show the ARPES data (for $\hbar \nu = 24$ eV) along $\bar{K}-\bar{\Gamma}-\bar{K}$ and $\bar{M}-\bar{\Gamma}-\bar{M}$ (for $\hbar \nu = 30$ eV) directions of the (001) surface BZ, respectively. ARPES results are consistent with the DFT predictions in panels (e) and (f).
To match the ARPES results with DFT, we have used a surface potential of $-0.14$~eV.
\label{fig3}}
\end{figure*}

Our spin-resolved ARPES measurements explicitly demonstrate the existence of a pair of type-II Dirac nodes in NiTe$_2$ along the $C_3$ rotation axis, lying just above (within 20 meV) the Fermi energy. Additionally, we show that NiTe$_2$ also hosts a series of inverted band-gaps (IBG). Especially, one of the IBG below the Fermi level supports a Dirac cone in the surface states. 
Together, the bulk Dirac node and the pair of IBG in NiTe$_2$ give rise to topological spin-polarized surface states over a wide range of energies. This non-trivial band morphology in NiTe$_2$ 
originates primarily from the 5$p$-orbital manifold of the Te atoms modified by the intra-layer hybridization, trigonal crystal field splitting and spin-orbit coupling. 

	

Bulk NiTe$_2$ crystallizes in the CdI$_2$ type trigonal structure (space group $P\bar{3}m1$, number 164). It has a layered structure with individual monolayer stacked together via weak van der Walls force. As shown in Fig.~\ref{fig1}(a)-(b), each monolayer has three sub-layers, with the central Ni layer being sandwiched between two adjacent Te layers (Ni-Te bond length 2.60~\AA). 
The observation of sharp spots in the low-energy diffraction pattern (LEED) in Fig.~\ref{fig1}(c) confirms the high quality of the NiTe$_2$ crystals cleaved along the (001) direction, along with the  presence of six-fold symmetry. Surface cleanliness of the as-cleaved samples was checked by high-resolution electron energy loss spectroscopy and X-ray photoelectron spectroscopy. The details of crystal preparation and characterization, ARPES measurements and DFT calculations are presented in Sec. S1, S2 and S3, of the Supplementary material (SM) 
\footnote{\label{SM} Supplementary material detailing 1) Sample preparation and characterization, 2) computational details, 3) ARPES measurements, 4) Topological band inversions and 5) ARPES measurements with potassium doping, is available at \url{https://bit.ly/2YWZf1o}}.

The electronic band structure of NiTe$_2$ including spin-orbit coupling is shown in Fig.~\ref{fig2}(a). It clearly depicts the presence of a pair of tilted band crossings along the $\Gamma-A$ direction. The presence of inversion and time-reversal symmetry mandates these bands to be doubly degenerate.
Furthermore, the $\Gamma-A$ high-symmetry direction is the invariant subspace of the three fold rotation ($C_{3}$) symmetry, and a symmetry analysis reveals that the crossing bands have opposite rotation character. This prevents their hybridization, resulting in a pair of gapless quadruply degenerate type-II Dirac points. 
The type-II nature of the DSM phase is also confirmed by the fact that the Dirac point appears at the touching point of the electron and hole pocket, as highlighted in Fig.~\ref{fig2}(b). 

Our photon-energy dependent ARPES data, in part presented in Fig.~\ref{fig2}(d-h), results in the $k_z$ dispersion presented in Fig.~\ref{fig2}(c). 
Our experimental results are consistent with the DFT-based bulk band structure calculations. Extrapolating the  fitted  DFT  band structure, we find that the  Dirac  cone is located  just  above  ($\sim$20 meV)  the  Fermi  energy. 
in contrast to other TMD-based type-II DSMs like PtTe$_2$\cite{PtTe2_natcomm,Bahramy2017,PtTe2PRL}, PtSe$_2$\cite{PtSe2_PRB}, and PdTe$_2$\cite{PdTe2_PRL,Bahramy2017,PhysRevB.96.041201}. 
Our attempt to electronically dope the sample via alkali metal (potassium) deposition (see Fig. S6 in the SM \cite{Note1}), and shift the Fermi energy above the Dirac point, revealed that only the surface states in NiTe$_2$ are impacted by surface deposition. Bulk doing is needed to shift the bulk bands. 


Extended energy range ARPES spectra for the two high-symmetry directions are shown in Fig.~\ref{fig3}(c) and (d). These spectra are mainly dominated by the surface states, as seen from the comparison with the calculated band structure in Fig.~\ref{fig3} (e) and (f). 
In order to understand their origin, we note that there are two symmetry inequivalent Te atoms (Te$^1$ and Te$^2$) and a single Ni atom in an unit cell of NiTe$_2$. The electronic configuration of Ni is $3s^23d^8$ and that of Te is $4d^{10} 5p^4$. We find that similar to other group-X TMDs \cite{Bahramy2017,Clark_2019}, the Te $5p$ orbital manifold in NiTe$_2$, aided by the 
interplay between intra-layer hopping, crystal field splitting, and SOC strength, gives rise to most of the bulk Dirac nodes and multiple inverted band-gaps. To highlight this, we show the evolution of the $p$-orbital manifold of the Te atoms in Fig.~\ref{fig3}(a). To start with (step I in Fig.~\ref{fig3}(a)), strong intra-layer hybridization between the Te$^1$ and Te$^2$ $p$ orbitals results in bonding and anti-bonding states. These orbitals are further split (in step II), due to a strong trigonal crystal field generated by the layered crystal structure of NiTe$_2$, separating $p_z$ from the $p_x,p_y$ orbitals. 
Inclusion of SOC (step III) further splits the orbitals into $|J,|m_J| \rangle$ states. Step IV of Fig.~\ref{fig3}(a) highlights the effect of the dispersion along the $\Gamma-A$ direction, and the formation of the bulk type-II Dirac point along with multiple band-inversions in the valance band. 

\begin{figure*}
\includegraphics[width=0.99\linewidth]{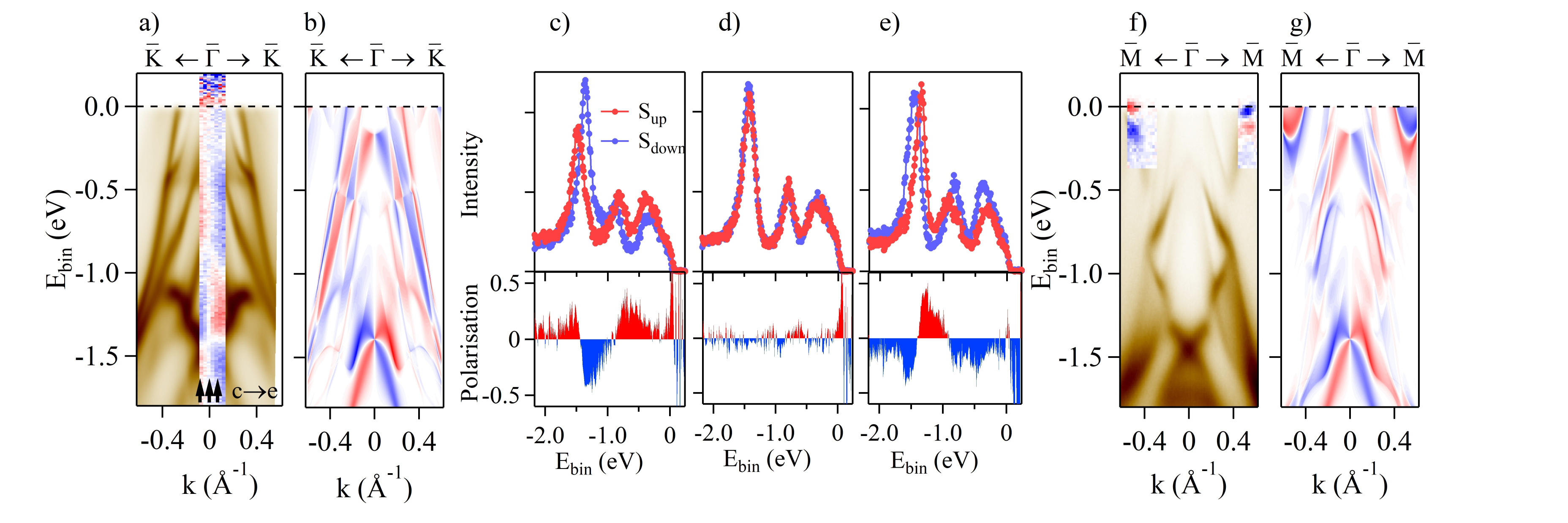}
\caption{Measured and calculated spin texture for the bands along $\bar{\Gamma}-\bar{K}$ (a, b) and $\bar{\Gamma}-\bar{M}$ (f, g) in the BZ. Experiments were performed in the same conditions as in Figure 3; (c-e) spin-polarized spectra and spin polarization for the points along $\bar{\Gamma}-\bar{K}$ marked by black arrows in (a); in all figures red/blue indicate positive/negative polarization perpendicular to the high-symmetry direction. $E_{bin}$ denotes the binding energy. 
\label{fig4}}
\end{figure*}

The irreducible representation of some of these states at the at $\Gamma$ and $A$ points and along the $\Gamma-A$ high-symmetry line is shown in Fig.~\ref{fig3}(b).
The bulk Dirac point originates from the crossing of the $\Delta_4$ and $\Delta_{5,6}$ states along the $\Gamma-A$ direction. As discussed earlier, the doubly degenerate $\Delta_4$ and $\Delta_{5,6}$ bands have  opposite rotation characters ($+1$ and $-1$, respectively) and, therefore, the Dirac point is protected from gap opening by any perturbation which respects the $C_3$ symmetry. Additionally, Fig.~\ref{fig3}(b) also highlights the existence of a pair of IBGs in the valance band at the $A$ point. 
However, in comparison to PdTe$_2$, the parity of the crossing bands at the $A$ point for NiTe$_2$ is different and only the lower IBG supports a Dirac node in the surface states. 
See Sec.~S5 in the SM \cite{Note1} for a more detailed discussion and comparison of the topological band structure and surface states with PdTe$_2$. 

%
%

The Dirac-like conical crossing in the surface states of NiTe$_2$ (at -1.4eV) is evident in the ARPES data taken along the two high-symmetry directions $\bar{K}-\bar{\Gamma}-\bar{K}$ and $\bar{M}-\bar{\Gamma}-\bar{M}$ as shown in Figs.~\ref{fig3}(c) and (d).  
The dominant surface bands are indicated by the red arrows. 
In addition to the Dirac cone, several other surface states are present in NiTe$_2$, owing to several band inversions below and above the Fermi level. Along the $\bar{M}-\bar{\Gamma}-\bar{M}$ direction, the surface states near the Fermi energy has its origin from a band inversion above the Fermi energy (see Fig.~\ref{fig3}(b), and Fig.~S3 in SM \cite{Note1}) and the corresponding surface Dirac cone lies above the Fermi level. While its Dirac-like nature is significantly altered far away from the $\bar{\Gamma}$ point, we demonstrate  its topological origin by displaying its chiral spin texture in Fig.~\ref{fig4}. Similar surface states have also been observed in other Te-based TMDs like PtTe$_2$ and PdTe$_2$, while they are absent in the Se based compounds like PtSe$_2$. This is a consequence of the avoided band inversion in PtSe$_2$ resulting from the reduced interlayer hopping \cite{PtTe2_natcomm,PhysRevLett.120.156401}.  

 
%


Since these surface states  have a topological origin, we now focus on the spin polarization of the bands using spin-polarized ARPES. 
In Fig.~\ref{fig4}(a) we display the spin-resolved data superimposed directly onto the spin-integrated band structure shown earlier in Fig.~\ref{fig3}(c) and (d). The measured spin polarization matches reasonably well with the calculated spin textures reported in Fig.~\ref{fig4}(b) and (g). In the present dataset, the spin component is always perpendicular to the dispersion direction. 
The most prominent feature in Fig.~\ref{fig4}(a)-(b) is the crossover of two opposite spin polarizations of almost equal magnitude for the surface state bands crossing at the $\Gamma$ point at a binding energy of $\sim -1.4$ eV. This confirms the helical nature of the spin-momentum locking in vicinity of the surface Dirac point, resulting from the IBG with the $Z_2 =1$ topological order. In Fig.~\ref{fig4}(c)-(e), we display the spin-polarized spectra and the spin polarization for the points marked by black arrows in Fig.~\ref{fig4}(a). The measured polarization perpendicular to $\bar{\Gamma}-\bar{K}$ reaches almost $50\%$. 

In the case of the bands along $\bar{M}-\bar{\Gamma}-\bar{M}$ direction of the surface BZ [Fig.~\ref{fig4} (f) and (g)], the polarization was measured for the electron-pocket-like surface states close to the Fermi energy. As discussed previously, although their shape is considerably different than a usual topological surface states, the clear spin polarization demonstrates their topological origin (see Fig. S3 of SM \cite{Note1}). These indeed appear to be the most prominent spin-polarized features also in the calculated spin polarization  perpendicular to the $\bar{\Gamma}-\bar{M}$ direction. The high values of the measured and calculated spin polarization indicates that NiTe$_2$ belongs to the recently identified topological-ladder family of Pt/PdTe2 \cite{Bahramy2017,Clark_2019}.

To summarize, we have established the existence of type-II DSM phase in NiTe$_2$ single crystals using spin-resolved ARPES measurements in conjunction  
with DFT-based {\it ab-initio} calculations. We show that, in contrast to similar class of materials like PtTe$_2$, PdTe$_2$, and PtSe$_2$, where the Dirac point is buried deep in the valence band, the Dirac point in NiTe$_2$ is located in vicinity of the Fermi energy. 
In addition to the bulk Dirac node, the Te $p$-orbital manifold in NiTe$_2$ also gives rise to a series of IBGs with non-trivial $Z_2$ topological orders. Together, these give rise to topological Dirac nodes in the surface states characterized by the particular spin texture over a wide range of energies. 
Our findings establish NiTe$_2$ as a prime candidate for exploration of Dirac fermiology and applications in TMD-based spintronic devices and ultrafast optoelectronics.

\section{acknowledgements}
A.A. acknowledges funding support by Dept. of Sci- ence and Technology, Government of India, via DST Grant No. DST/NM/NS/2018/103(G), and from SERB Grant No. CRG/2018/002440.
B.G. acknowledges CSIR for senior research fellowship. 
A. A. and B. G. acknowledges HPC- IIT Kanpur for its computational facilities. This work has been partly performed in the framework of the nanoscience foundry and fine analysis (NFFA-MIUR Italy, Progetti Internazionali) facility.

{\it Note:} B. G. and D. M. contributed equally to this work as first authors. 

\bibliography{NiTe2}

\end{document}